\documentclass[oldversion]{aa}  

\newcommand{\lens}{SL2S\,J08544-0121}

\usepackage{graphicx}
\usepackage{txfonts}
\usepackage{natbib}
\begin{document}
 
\title{
Strong Lensing as a Probe of the Mass Distribution \\ \emph{Beyond} the Einstein Radius}
   \titlerunning{Strong Lensing Beyond the Einstein Radius}
   \authorrunning{Limousin et~al.}
   \subtitle{Mass \& Light in \lens, a Galaxy Group at $z=0.35$}
   \author{Marceau Limousin\inst{1,2,3}, Eric Jullo\inst{4,1},
   Johan Richard\inst{5,6}, R\'emi Cabanac\inst{2},\\ Sherry H. Suyu\inst{7}, 
Aleksi Halkola\inst{8}, Jean-Paul Kneib\inst{1},
   Raphael Gavazzi\inst{9,10} \& Genevi\`eve Soucail\inst{11}
      \thanks{Based on observations obtained with MegaPrime/MegaCam, a joint
       project of CFHT and CEA/DAPNIA, at the Canada-France-Hawaii Telescope
       (CFHT) which is operated by the National Research Council (NRC) of Canada,
       the Institut National des Sciences de l'Univers of the Centre National
       de la Recherche Scientifique (CNRS) of France, and the University of
       Hawaii. This work is based in part on data products produced at TERAPIX
       and the Canadian Astronomy Data Centre as part of the Canada-France-Hawaii
       Telescope Legacy Survey, a collaborative project of NRC and CNRS.
       Also based on HST data, program 10876 and Keck telescope data.}
       }
   \offprints{marceau.limousin@oamp.fr}

   \institute{
	Laboratoire d'Astrophysique de Marseille, UMR\,6610, CNRS-Universit\'e de Provence,
	38 rue Fr\'ed\'eric Joliot-Curie, 13\,388 Marseille Cedex 13, France
	\and
	Laboratoire d'Astrophysique de Toulouse-Tarbes, Universit\'e de Toulouse, CNRS,
	57 avenue d'Azereix, 65\,000 Tarbes, France
   	\and
         Dark Cosmology Centre, Niels Bohr Institute, University of Copenhagen,
        Juliane Maries Vej 30, 2100 Copenhagen, Denmark
	\and
	Jet Propulsion Laboratory, California Institute of Technology, Pasadena, CA 91109, USA
	\and
	Durham University, Physics and Astronomy Department, South Road, Durham DH3 1LE, UK
	\and
	Department of Astronomy, California Institute of Technology, 105-24, Pasadena, CA91125, USA
	\and
	Argelander-Institut f\"ur Astronomie, Universit\"at Bonn, Auf dem H\"ugel 71, 53121 Bonn, Germany
	\and
	Excellence Cluster Universe, Technische Universit\"at M\"unchen, Boltzmannstr. 2, 85748 Garching, Germany
	\and
         CNRS, UMR\,7095, Institut d'Astrophysique de Paris, F-75014, Paris, France
	\and
	 UPMC Universit\'e Paris 06, UMR\,7095, Institut d'Astrophysique de Paris, F-75014, Paris, France
	\and
	Laboratoire d'Astrophysique de Toulouse-Tarbes, Universit\'e de Toulouse, CNRS,
	14 avenue Edouard Belin, 31\,400 Toulouse, France
              }

   
  \abstract
   {
   Strong lensing has been employed extensively to obtain accurate
   mass measurements \emph{within} the Einstein radius.  In this
   article, we use strong lensing to probe mass distributions
   \emph{beyond} the Einstein radius.
   We consider \lens, a galaxy group at redshift
   $z=0.35$ with a bimodal light distribution and with a strong
   lensing system located at one of the two luminosity peaks 
   separated by  $\sim$$\,54\,\arcsec$. The main arc and the counter-image of
   the strong lensing system are located
   at $\sim$\,5$\arcsec$ and $\sim$\,8$\arcsec$, respectively, from
   the lens galaxy 
   centre. We find that a simple elliptical isothermal potential
   cannot satisfactorily reproduce the strong lensing observations.
   However, with a mass model for the group built from its
   light-distribution with a smoothing factor $s$ and a
   mass-to-light ratio M/L, we obtain an accurate reproduction of the
   observations. We find M/L\,=\,98\,$\pm$\,27 ($i$ band, solar units,
   not corrected for evolution) and $s$\,=\,20$\arcsec \pm$\,9 (2
   $\sigma$ confidence level). Moreover, we use weak lensing to
   estimate independently the mass of the group, and find a consistent
   M/L in the range 66-146 (1-$\sigma$ confidence level).  This
   suggests that light is a good tracer of mass. Interestingly, this
   also shows that a strong lensing \emph{only} analysis (on scales of
   $\sim$ 10$\arcsec$) can constrain the properties of nearby objects
   (on scales of $\sim$ 100$\arcsec$). We characterise the type of
   perturbed strong lensing system allowing such an analysis: {\emph{a
   non dominant strong lensing system used as a test particle
   to probe the main potential.}} This kind of analysis needs to be validated
   with other systems since it could provide a quick way of probing the
   mass distribution of clusters and groups. This is particularly relevant in the
   context of forthcoming wide field surveys, which will yield
   thousands of strong lenses, some of them being perturbed
   enough to pursue the analysis proposed in this paper.}


   \maketitle
 
\section{Introduction}

Gravitational lensing probes the mass distribution projected
    along the line of sight.
When the surface mass density of a lens is larger than a critical
threshold, i.e. in the strong lensing (SL) regime, the light from a
background source galaxy is lensed into multiple images.  
These multiple images provide strong observational constraints on the projected
mass distribution of the lens within the Einstein radius.
Since the discovery of the first gravitational arc in the galaxy
cluster Abell~370 twenty years ago \citep{lynds1,soucail87,a370}, strong
lensing has been widely used to probe the mass distribution
of structures at different scales: galaxies \citep[see, \emph{e.g.} the SLACS survey,][]{slacs3},
galaxy clusters \citep[see, \emph{e.g.}][]{halkola} and recently galaxy groups \citep{sl2s,paperI,belokurov08}.

\subsection{Perturbing a Strong Lensing System}
Because most of the galaxies in the Universe are part of larger structures, either groups or
clusters, so are many SL systems
\citep[see, \emph{e.g.}][]{kundic97a,fassnacht02,cecile04,morgan05,williams06,momcheva,auger07,ring1689,auger08,grillo12,treuaroundslacs,inada}.
A mass distribution located at a small angular distance from a strong lens may induce measurable perturbations in the 
lensing signal. Not taking this external perturbation into account can seriously bias the results inferred from the
SL modelling as shown by \citet{keeton04}:
they found that if the environment is neglected, SL modelling of double-image lenses 
largely overestimate both the ellipticity of the lens galaxy ($\Delta e/e \sim 0.5$) and the Hubble constant
($\Delta h/h \sim 0.22$). 
In contrast, models of four-image lenses in which perturbations by the
environment are incorporated as a tidal shear,
 recover the ellipticity reasonably well, while still overestimating the
Hubble constant ($\Delta h/h \sim 0.15$). They argue that most of the
biases are due to the neglected convergence from nearby massive groups or clusters of galaxies.
More generally, the topic of modelling a lens with an external mass perturbation has been addressed by 
different authors \citep[see, \emph{e.g}][and references therein]{keeton97,ReImportance,keeton04,oguri05,oguri08,dye07}. 

To summarise, a precise SL modelling can be \emph{affected} by external mass 
distributions, and people have tried to take this bias into account in
order to improve the SL modelling. 
As observations become more and more accurate, we can expect to be more and more
sensitive to external mass distributions near strong lenses.
In this article, we propose to 
\emph{exploit this external effect} by using
the perturbations measured in SL modelling as \emph{probes of the external mass distribution}.

\subsection{The ``Ring'' Test in Abell~1689}
We first remind the reader of a previous attempt we made to locally probe 
the potential of the galaxy cluster Abell~1689 with a perturbed SL system.
In the core of galaxy cluster Abell~1689, \citet{mypaperIII} reported
SL systems (``rings'') formed around three elliptical galaxies
located 100$\arcsec$ away from the cluster centre, i.e. the
transitional region between the strong and weak lensing
regimes.
These SL systems should be sensitive to the external shear and convergence produced
by their parent cluster \citep{KochanekBlandford91}.
Based on simulations, \citet{ring1689} showed that such strong lenses could be
used to probe the cluster potential locally. They applied this method to the
three rings discovered in Abell~1689, and found that solely modelling
these three rings 
(i.e. without including any other multiply-imaged systems that are also produced by the cluster) 
provides strong evidence for bimodality of the cluster core; it is not possible to model 
simultaneously the three rings assuming a unimodal mass distribution for the cluster.
This bimodality confirms previous parametric SL studies of Abell~1689 
\citep{miraldababul95,halkola,mypaperIII,1689flexion,saha07,japflexion}.
More importantly, this result shows that SL features of 1-2$\arcsec$-wide Einstein rings
actually contain information on the mass distribution of the parent cluster, i.e. on a much larger 
scale than their Einstein radii. In other words, this study suggests that strong lenses can be used 
to probe mass distributions \emph{beyond} their Einstein radius.
In this article, we further develop this idea on another perturbed SL system located in a galaxy group, \lens.

All results are scaled to a flat, $\Lambda$CDM cosmology with $\Omega_{\rm M}$=0.3,
$\Omega_{\Lambda}$=0.7 and a Hubble constant H$_0$=70 km\,s$^{-1}$\,Mpc$^{-1}$. 
In this cosmology, 1$\arcsec$ corresponds to a physical transverse
distance of 4.94\,kpc at $z=0.35$.
All images are aligned with the WCS coordinates, i.e. north is up and east is left. Magnitudes are given in the
AB system. Luminosities are given for the $i$ band, in solar units, not corrected for passive evolution.
Ellipticities are expressed as $(a^2-b^2)/(a^2+b^2)$, and position angles
are given counterclockwise with respect to the west.
Shear and convergence are computed for a source redshift of $z_s$=1.268.

\section{\lens: Presentation \& Data}
\lens\, is part of the Strong Lensing Legacy Survey \citep[SL2S,][]{sl2s} ,
which collects SL systems in the Canada France Hawaii
Telescope Legacy Survey (CFHTLS).
\lens\, is a galaxy group at $z=0.35$ presented by \citet{paperI} which contains a SL system
(Fig.~\ref{present}).

\subsection{Ground-Based Imaging}
\lens\, has been observed in five bands as part of the
CFHTLS. The $i$-band data are used to build luminosity maps from isophotal 
magnitudes of elliptical group members and to perform a weak-lensing analysis.

The bottom panel of Fig.~\ref{present} shows a 10'\,$\times$\,10' CFHTLS $i$-band image. The white cross 
gives the location of the strong lens. We draw luminosity isodensity
contours of 10$^{5}$, 3$\times$10$^{5}$, 10$^{6}$, 3$\times$10$^{6}$ and 10$^{7}$ L$_{\sun}$\,kpc$^{-2}$. 
The top-right panel also shows a CFHTLS 1-arcmin$^2$  $gri$ colour image
centred on the lens.

\subsection{Space-Based Imaging}
\label{space}
The strong lensing features detected from ground-based images have been observed with the
\emph{Hubble Space Telescope} (HST). Observations were done in snapshot mode (C15, P.I. Kneib,
ID 10876) in three bands with the ACS camera (F475, F606, and F814).
Figure~\ref{present} shows a colour image of the strong lens based on these observations. We report two 
multiply-imaged systems: the first system is bright and forms a typical cusp configuration perturbed by a 
satellite galaxy (labelled Dwarf on Fig.~\ref{present}).
The second system is a very faint arc located west of the lens at a larger radius.
It is not possible to reliably identify individual images on
the faint arc.
Moreover, given its faintness, spectroscopy is hopeless 
with current facilities, as the surface brightness is ca. 31 mag\,arcsec$^{-2}$, and therefore 
it is not used in the following analysis.
As can be appreciated on Fig.~\ref{present}, the HST data brings
significant amounts of
additional information on the lensed features.

\subsection{Spectroscopy}
\begin{figure}[h!]
\begin{center}
\includegraphics[scale=0.3]{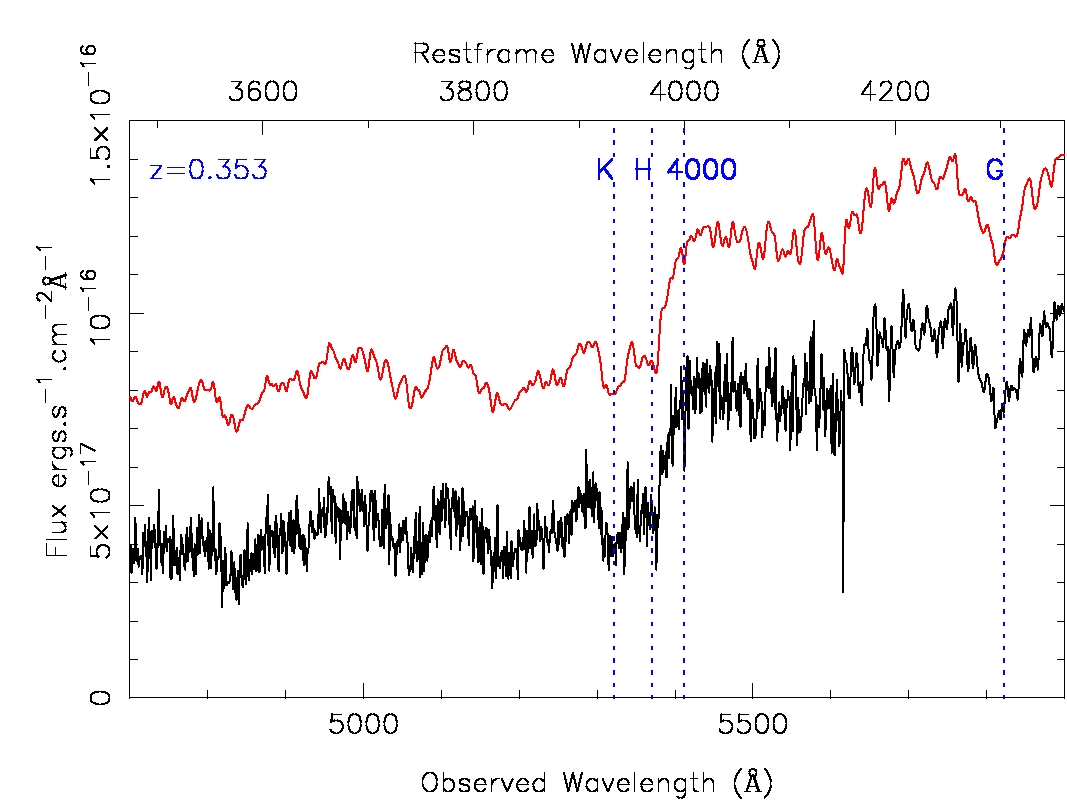}
\includegraphics[scale=0.3]{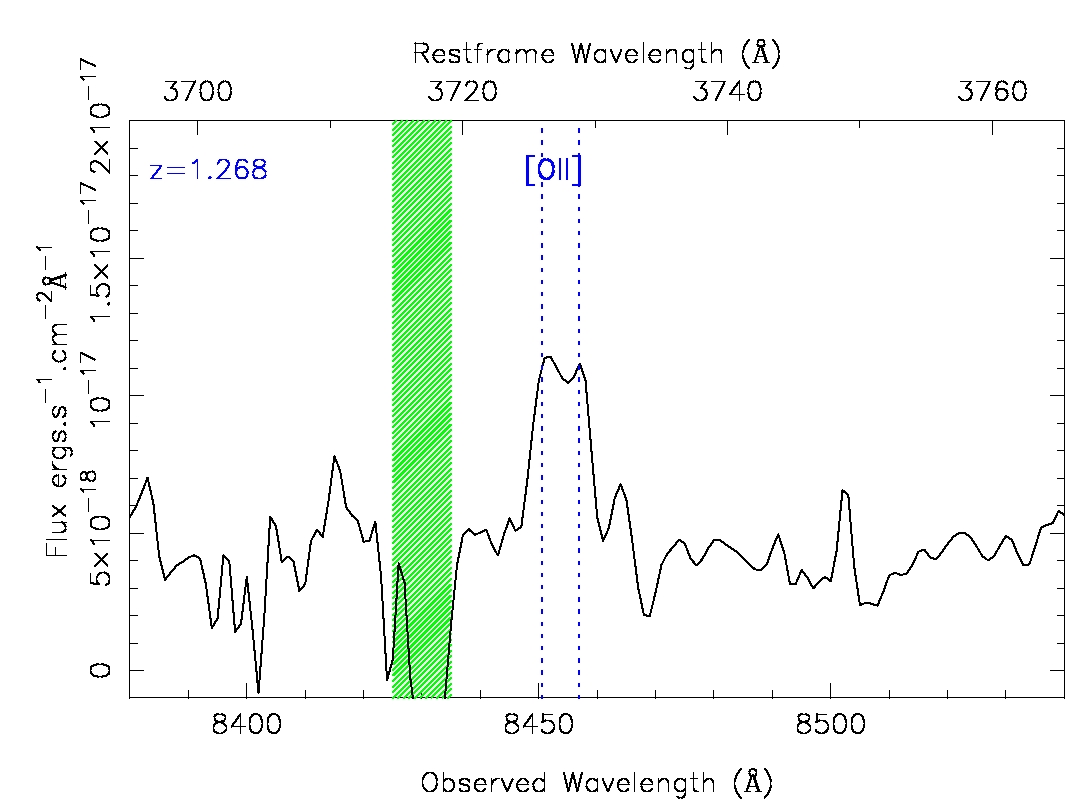}
\caption{
1-D spectra of the lens (\emph{up}) and of the bright arc (\emph{low}).
The green shaded region masks the residuals of a strong OH atmospheric emission
}
\label{spectro}
\end{center}
\end{figure}

We have used the Low Resolution Imager and Spectrograph \citep[LRIS,][]{lris} on the Keck telescope to 
measure the spectroscopic redshift of both the lens and the brightest arc of the \lens\, system. 
On January 14 2007, we obtained 300 seconds on the lensing galaxy and 4 exposures 
of 900 seconds each
on the arc, using a 1.0\arcsec\ wide slit. A 600 lines mm$^{-1}$ grism blazed at 4000 \AA\ and a 400 lines 
mm$^{-1}$ grating blazed at 8500 \AA\ were used in the blue and red channels of the instrument,
both light paths being separated by a dichroic at 5600 \AA. The corresponding dispersions are 0.6/1.85 \AA\ 
and resolutions are 4.0/6.5 \AA\ in the blue/red channel. The resulting extracted spectra are shown in
Fig.~\ref{spectro}. The lens presents a typical elliptical spectrum at $z=0.3530\pm0.0005$ with prominent 
H and K CaII lines, 4000 \AA\  break,  and G band. 
The spectrum of the arc shows a strong emission line at 8454 \AA, resolved in a doublet separated by $\sim5$ \AA\ . 
This gives us an unambiguous redshift determination of $z=1.2680\pm0.0003$ with [OII] emission.

\subsection{Global Properties}
\begin{figure*} \begin{center}
\includegraphics[width=0.918\columnwidth]{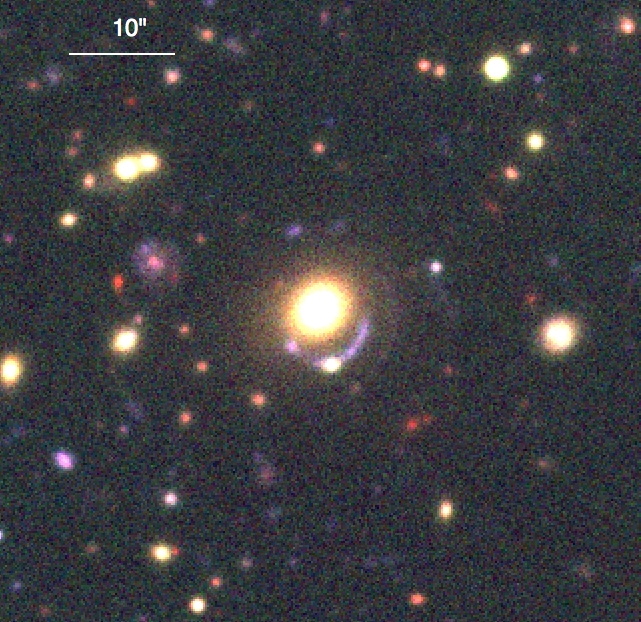}
\includegraphics[width=0.918\columnwidth]{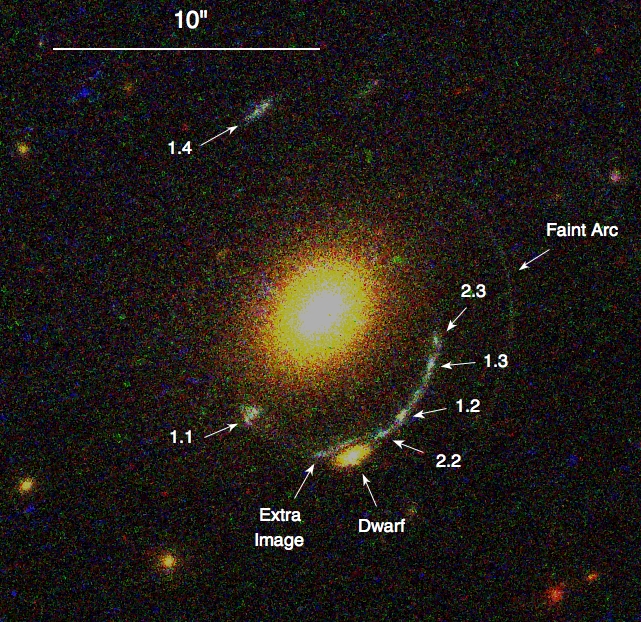}
\includegraphics[width=1.9\columnwidth]{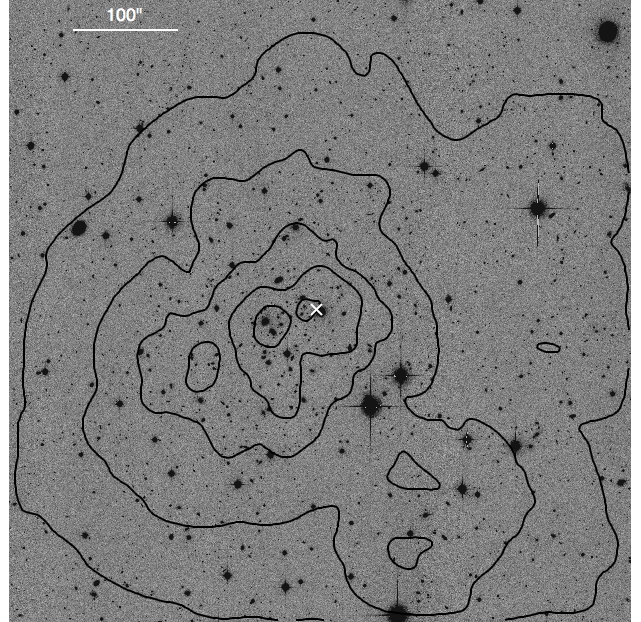}
\caption{Group SL2S\,J08544-0121 at $z_{\rm spec}=0.353$.
\emph{Upper Left:} composite CFHTLS $gri$ colour image
(1 arcmin$^2$ = 297 $\times$ 297 kpc$^{2}$).
\emph{Upper Right:} composite HST/ACS F814W-F606W-F475W colour image 
($24\arcsec \times 24\arcsec$ = 118 $\times$ 118 kpc$^{2}$).
We show the proposed multiple-image identification. The dwarf
galaxy and the main extra image it produces is labelled.
\emph{Lower:} CFHTLS $i$ band ($10'\,\times\,10'$ = 2\,969 $\times$ 2\,969 kpc$^{2}$).
Luminosity isodensity contours of
10$^{5}$, 3\,10$^{5}$, 10$^{6}$, 3\,10$^{6}$ and 10$^{7}$ L$_{\sun}$\,kpc$^{-2}$ are drawn (continuous black line), and the white cross shows the location of the SL system.}
\label{present} \end{center} \end{figure*}

The luminosity contours of \lens\, are elongated in the east-west direction.
The SL deflector is populated by a single bright galaxy.
Note (Fig.~\ref{present}) that the innermost luminosity isodensity contour at
10$^{7}$ L$_{\sun}$\,kpc$^{-2}$  encompasses the SL system but also two bright galaxies located
$\sim$ 54$\arcsec$ east from the SL system, making this light distribution bimodal.
This is the only group of the sample presented by \citet{paperI} for which the luminosity
isodensity distribution is not clearly dominated by the lens, making this configuration rather exceptional:
the large Einstein radius ($\sim5\arcsec$) indicates a significant mass concentration 
associated with this lens, but the luminosity isodensity distribution is actually bimodal.

\section{Modelling the Lens}
In this Section, we attempt to reproduce the SL multiple images using a single elliptical isothermal
potential centred on the bright galaxy.
All optimisations are performed in the \emph{image plane} using the 
\textsc{Lenstool} software \citep{jullo07}. 
We quantify the goodness of the fit by using the image plane RMS and 
the corresponding $\chi^2$. When necessary, we compare the fits using the Bayesian Evidence.

\subsection{Observational Constraints}
As explained in Section~\ref{space}, we do not use the faint arc in the analysis and
focus on the bright multiply-imaged system.
This system is composed of 4 main images: 1.1, 1.2, 1.3 and 1.4.
Additional images are produced by the satellite (dwarf) galaxy. These are not considered in the
analysis because we do not want to complicate the modelling by adding a sub-halo for the
satellite galaxy. We also \emph{make the assumption} that, given its small size, 
the satellite galaxy does not influence images 1.1 to 1.4. This assumption will be discussed in 
Section~\ref{discuss} and addressed further in a forthcoming publication.
Since the merging arc composed of images 1.2 and 1.3 is well resolved, we can safely associate
two other images on this arc, namely 2.2 and 2.3 (Fig.~\ref{present}).
Their counter-images expected near images 1.1 and 1.4 are not safely identified, therefore
we do not use them in the analysis.
Indeed, parametric strong lensing analyses are highly sensitive to
misidentifications of images
and we prefer to use only the images we are confident in. This gives us a total of 8 observational constraints.

\subsection{Shape of the Bright Galaxy}
\label{PAlight}
In this subsection, we describe the properties of the light distribution of the bright galaxy populating
the strong lensing deflector.
We use the IRAF task \emph{ellipse} to measure the shape of its isophotes.
We find an ellipticity of $e=0.206$ and a position angle of 39$\pm$5 degrees at 
2$\arcsec$ from the centre.
A closer inspection of the galaxy centre clearly reveals a double core even though the 
outer isophotes are elliptical. The above measurements therefore correspond to the 
superposition of the light from each component. The spectrum of the galaxy presented in 
Fig.~\ref{spectro} does not show features of another galaxy at a different lens redshift along 
the line of sight. Given the similar colours of these two components, the bright galaxy 
may be the result of a recent merger.

\subsection{Positional Uncertainty}
The sizes of the multiple images have been estimated using the IRAF task \emph{imexamine}.
They range from 0.11$\arcsec$ to 0.15$\arcsec$, with a mean of 0.13$\arcsec$.
Therefore, the positional uncertainty is set to 0.13".
We note that this positional uncertainty may seem large for HST data but we stress that we are not 
in the case of point-like objects (quasar lensing) where the astrometric precision can reach 0.01$\arcsec$. 
In our case, the images are extended and the depth of the snapshot observations does not allow us
to resolve better the conjugated points with \emph{imexamine}.
We note that large positional uncertainties are often used in the case of extended
images \citep[see, \emph{e.g.}][]{oguri10}.

\subsection{Mass Model}
\label{singlemodelling}

The lens potential is parametrised by a dual Pseudo Isothermal Elliptical Mass Distribution 
\citep[dPIE, see][]{ardis2218}.
The 3D density distribution of the dPIE is: 
\begin{equation}
\rho(r)={\rho_0 \over (1 + {r^2 / {r^2_{\mathrm{core}}}}) (1 + {{r^2 / r^2_{\mathrm{cut}}}})}; \ \ \ \ r_{\mathrm{cut}}>r_{\mathrm{core}}.
\end{equation}
This distribution represents a spherical system with scale radius $r_{\mathrm{cut}}$, core radius $r_{\mathrm{core}}$ and central density $\rho_0$.

This profile is formally the same as the Pseudo Isothermal Mass Distribution (PIEMD) described in
\citet{mypaperI}.
Its scale radius is set to 250\,kpc, i.e. larger than the range where the observational constraints are found.
Allowing $r_{\mathrm{core}}$ to vary produces models with core radii much smaller
than the range of radii over which we have observational constraints. Therefore, we can set 
$r_{\mathrm{core}}$\,=\,0,
and the dPIE profile are close to isothermal in the range of interest. The remaining free parameters 
of the dPIE profile are:
\begin{itemize}
\item[] - the halo centre position (X,Y), which is allowed to vary within 3$\arcsec$ of the  light distribution 
centre
\item[] - the halo ellipticity $e$, which is forced to be smaller than 0.6, as suggested by numerical simulations \citep{jingsuto}
\item[] - its position angle $\theta$, which is allowed to vary between 0 and 180 degrees
\item[] - The fiducial velocity dispersion{\footnote{linked to the central density by: 
$\sigma_0^2=\frac{4}{3}G\pi \rho_0 
\frac{r_{\mathrm{core}}^2 r_{\mathrm{cut}}^2}{(r_{\mathrm{cut}}-r_{\mathrm{core}})(r_{\mathrm{cut}}+r_{\mathrm{core}})^2},
$
}}
which is allowed to vary between 200 and 900 km/s.
\end{itemize}
We emphasise that this fiducial velocity dispersion is not the
Spherical Isothermal Sphere velocity dispersion. It is usually
smaller, and we refer the reader to \cite{ardis2218} for a self-contained description of the dPIE profile.

\subsection{Results: Bad Optimisation}
Results of the optimisation are given in Table~\ref{res}.
This first optimisation results in a poor fit to the data, with the RMS error of image positions $\sim0.38\arcsec$ in the image plane (i.e. 
significantly larger than the assumed positional uncertainty of 0.13$\arcsec$) and a reduced 
$\chi^2$ of 29. The halo position is found to coincide with the light distribution centre within error bars.
The halo ellipticity is at the upper bound of the input prior, and the position angle is equal to $\sim$\,18 deg. 
Only when we allow the halo ellipticity to reach values as high as 0.9 are we able to reproduce the 
observational constraints (RMS equals to 0.06$\arcsec$ for $e=0.9$, $\theta$\,$\sim$ 19.5 degrees, and the halo centre is offset from the light distribution centre by one arcsecond.

We conclude that a single potential does not satisfactorily reproduce the observational constraints.
We have used the lens modelling method based on
\citet{halkola,halkola08} in parallel to our
method, and found that the observational constraints used in this work
require an external shear component in order to be properly
reproduced.
In the rest of the paper, we include the contribution of the external mass distribution in the lens modelling.

\begin{table*}
\begin{center}
\caption{
Parameters of the lens inferred from two optimisations:
\emph{First line:} A single halo models the lens potential.
\emph{Second line:} We add an external mass perturbation on top of the halo lens potential.
Coordinates are given in arcseconds with respect to the centre of the galaxy deflector.
$e$ is the central halo ellipticity. Error bars correspond to $1\sigma$ confidence levels.}
\label{res}
\begin{tabular}{cccccccccc}
\hline
\hline
Model& $\delta(x)$ & $\delta(y)$ & $e$ & $\theta$ & $\sigma_0$ (\footnotesize{km\,s$^{-1}$}) & RMS & $\chi^2$ & log(Evidence) & Prior \\
\hline
1 Lens & -0.24$\pm$0.12 & -0.02$\pm$0.10 & 0.597$^{+0.002}_{-0.038}$ & 17.9$\pm$1.6 & 476$\pm$7 & 0.38$\arcsec$ & 86/3 & -62 & $e<$ 0.6\\
\hline
Ext. Perturb. & 0.01$\pm$0.05 & -0.26$\pm$0.08 & 0.499$\pm$0.04 & 21.5$\pm$0.9 & 447$\pm$3 & 0.05$\arcsec$ & 0.96 & -20 & $e<$ 0.6 \\
\hline
\hline
\end{tabular}
\end{center}
\end{table*}

\section{An External Mass Perturbation Based on the Light Distribution: Does Light traces Mass?}

The large scale properties of \lens\, shown on Fig.~\ref{present}, together with the 
failed modelling attempted in the previous Section, suggest the need to take into account an external mass
perturbation. In order to test the hypothesis that light traces mass, this external perturbation will be mapped
from the known light distribution properties.

\subsection{Luminosity Maps}
The first step is to build luminosity maps of \lens\, from which we will derive the external mass
perturbation properties.
To identify group members, we select all galaxies having a $r-i$ colour difference smaller than 0.15 
magnitudes from the bright galaxy deflector \citep{paperI}.
Because we want to describe the perturbation of the galaxy group \emph{on} the SL system,
this luminosity map should not take into account the light coming from the galaxies populating the SL 
deflector.
Therefore, we select all group members \emph{apart from the bright galaxy populating the deflector and the associated
satellite galaxy}. This partial group luminosity is referred to as L$_{\rm ext}$ hereafter.
From this catalogue, we generate smoothed luminosity maps, and hence the mass
maps, assuming mass follows light. An important ingredient of this
procedure is the smoothing scale of the luminosity maps.  Since this
influences the properties of the derived mass maps, we
adopt the smoothing scale as a free parameter  
for describing the external mass perturbation.

\begin{figure*}
\begin{center}
\includegraphics[scale=0.66]{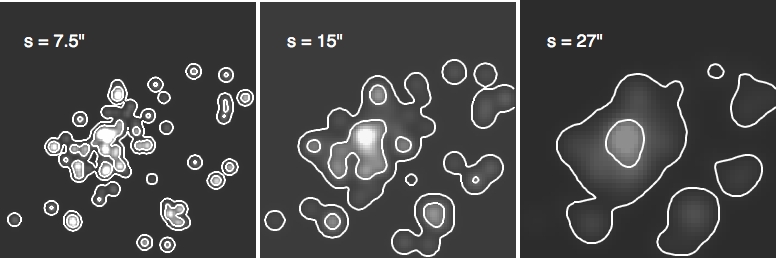}
\caption{
Three luminosity maps of the same luminosity L$_{\rm ext}$ for different smoothing scales as indicated on 
each panel. We draw in white luminosity isodensity contours of 
10$^{7}$ and 10$^{8}$ L$_{\sun}$\,arcsec$^{-2}$.
The smoothing scale $s$ influences the shape of the luminosity isodensity 
contours, and by construction the shape of the resulting mass distribution, hence its lensing properties.
}
\label{lumaps}
\end{center}
\end{figure*}

We use the following smoothing scheme:
the 10'\,$\times$\,10' CFHTLS $i$-band image is divided into cells of size $c$ pixels, which
translates into $c$\,$\times$\,0.186 $\arcsec$  (since the pixel size equals 0.186$\arcsec$). We compute the rest-frame $i$-band luminosity $L$ of each 
galaxy located in a given cell with
\begin{equation}
L = 10^{ ( M_{\sun} - M + \rm{DM} + k) / 2.5}
\end{equation}
where $M$ is the $i$-band isophotal magnitude of the galaxy, $M_{\sun}$ is the solar absolute magnitude in
the $i$ band, DM is the distance modulus, and $k$ the k-correction
factor that is estimated from elliptical 
templates by \citet{BC03} using single-burst stellar formation models. Then we sum up the luminosities of 
all galaxies in each cell to get the total luminosity for the cell. The resulting luminosity isodensity map is then 
convolved with a Gaussian kernel of width $w$. This gives an angular
smoothing scale $s$ that equals 
$c$\,$\times$\,0.186\,$\times\,w\,\arcsec$. 
Figure~\ref{lumaps} shows three luminosity maps, where we distribute the same total luminosity L$_{\rm ext}$ for three smoothing scales. We draw luminosity isodensity contours of 10$^{7}$ and 10$^{8}$ L$_{\sun}$\,arcsec$^{-2}$.
One can appreciate how the smoothing scale $s$ influences the shape of the luminosity isodensity contours.

\subsection{From Light Map to Mass Map}
Once a luminosity map with a given smoothing scale $s$ is obtained, we
assume a constant mass-to-light ratio 
M$_{\rm ext}$/L$_{\rm ext}$ to convert it into a mass map. 
This M$_{\rm ext}$/L$_{\rm ext}$ is the second free parameter describing the perturbation produced by the 
galaxy group.
Since we have excluded the bright galaxy populating the deflector 
when building the luminosity map, this mass map can be considered as the external mass perturbation.
Therefore we refer to the mass contained in this map as M$_{\rm ext}$. Then, we use the algorithm
developed by \citet{jullo09} to transform this mass map into a grid of analytic circular dPIE potentials, 
supported by \textsc{Lenstool}.

We model the mass distribution of SL2S J08544-0121 with a 5$'$ hexagonal grid of dPIE potentials. 
In order to build an adaptive grid where the resolution follows the 2-D mass density, we recursively split the 
input mass map into equilateral triangles until the  mean surface mass
density per triangle is lower than 
10$^7$ M$_{\sun}$\,arcsec$^{-2}$. Then we place a dPIE potential at each node of the grid with the 
following parameters: core radii $r_{\mathrm{core}}$ are set to the local grid resolution and cut-off radii  
$r_{\mathrm{cut}} =  3 \times r_{\mathrm{core}}$. In \citet{jullo09}, we found that such values of 
$r_{\mathrm{cut}}$ ensured a smooth and extended density profile. We estimate the dPIE central 
velocity dispersions $\sigma_i$ by inverting the equation 
\begin{equation}
\sigma_i^2 = M_{i,j} \Sigma_j,
\label{equ1}
\end{equation}
where $\Sigma_j$ is the surface mass density at the grid nodes location, and $M_{i,j}$ is a mapping matrix whose 
coefficients depend of the dPIE core and cut-off radii \citep[see][]{jullo09}. In order to prevent negative 
$\sigma_i^2$, we invert equation~\ref{equ1} by a mean-square minimisation technique.
The density threshold (for splitting the cells into triangles)
controls the grid resolution, and might thus be considered as an important 
parameter.  However, we have tried to use smaller thresholds, down to 10$^{5}$ M$_{\sun}$\,arcsec$^{-2}$, and the 
results were unchanged.  Therefore, we keep the 10$^{7}$ M$_{\sun}$\,arcsec$^{-2}$ threshold because
the corresponding mass maps require less mass clumps. We also force
the algorithm to stop after 4 levels of splitting. On average, a grid
cell contains about 200 dPIE potentials.

\subsection{Modelling the Lens including the External Perturbation}
We now model the SL system, taking into account the external mass perturbation 
parametrised by a smoothing scale $s$ and a mass-to-light ratio M$_{\rm ext}$/L$_{\rm ext}$.
We generated mass maps with smoothing scales $s$ ranging from 1 to 40$\arcsec$
in steps of 2.5$\arcsec$  and mass-to-light ratios  from 10 to 190 in steps of 20.
Each mass map is then included in the modelling of the SL system. This modelling is performed in the image 
plane. 
We note that these two extra parameters describing the external mass perturbation are not treated the
same way as the five parameters of the deflecting halo. For each set of parameters
($s$, M$_{\rm ext}$/L$_{\rm ext}$), we optimise
the 5 parameters of the halo.
Parameters for the strong lens deflector are the same as in Section~\ref{singlemodelling}. For each set of
parameters ($s$, M$_{\rm ext}$/L$_{\rm ext}$), we quantify the goodness of the SL modelling using the 
image plane RMS, the corresponding $\chi^2$ and the Bayesian Evidence.

Since our goal is to constrain the galaxy group as a whole, in the
following we use M/L corresponding to 
the \emph{total} mass-to-light ratio of the group; i.e. M (L) is the sum of the external mass (luminosity) 
perturbation \emph{and} the mass (luminosity) of the lens. 
We checked that degeneracies of each mass component near the lens are small. 
For the range of 
parameters ($s$, M$_{\rm ext}$/L$_{\rm ext}$) investigated in this work, we compute 
M$_{\rm ext}$/M$_{\rm lens}$ in a circle of radius 10$\arcsec$ centred on the lens.
This ratio falls between 10$^{-7}$ and 10$^{-5}$.

Total masses and luminosities are computed within a region of 10'\,$\times$\,10' centred on the lens.
At the redshift of the group, it corresponds to $\sim$\,3\,$\times$\,3 Mpc$^{2}$ (Fig.~\ref{present}).

\section{Results: Properties of \lens}
\label{result}
For a certain range of parameters characterising the external mass perturbation
we obtain excellent fits to the observed constraints.
We present first the best-fit model, and then the derived constraints on the  galaxy group properties.
We emphasise that what we achieve here is to constrain the properties of the galaxy group as a 
\emph{whole}  (on scales of 100$\arcsec$) based on a local SL analysis \emph{only} (on scales of 
10$\arcsec$).

\subsection{Best-Fit Model for the Lens}
The modelling results are given in Table~\ref{res}. The best-fit model
has a total mass-to-light ratio of $\simeq 75$
($i$ band, solar units, not corrected for evolution) and a smoothing
scale of $\simeq 20\arcsec$ 
(Fig.~\ref{results}). The RMS error between observed and modelled image positions in the image plane is 
0.05$\arcsec$, yielding a reduced $\chi^2$ of 0.96. This is a significant improvement compared to the 
modelling without external mass perturbation, which had
RMS\,=\,0.38$\arcsec$.  
To compare quantitatively the two models  (i.e. mass models with and without the
inclusion of external perturbations), we compute the Bayesian evidence values of
the two models.  The evidence takes into account the additional complexity of
the new model with the extra parameters for the external
perturbations.  The difference in the evidence of the two models,
which is the relative probability of the models given the data
(assuming the two models are equally probable a priori), is
$2\times 10^{18}$. The data therefore rank the perturbed model much higher
than the simple model.

We find the position of the halo to coincide with the centre of the light distribution.
The modelled position angle of the halo is 21.5 deg. Comparing this value to the position angle of the 
light distribution is complicated due to the bimodal light distribution of the bright galaxy (Section~\ref{PAlight}).
In particular, a merger will affect the light and mass distributions so that agreement may not necessarily be expected.

\subsection{Constraints on the Group as a Whole}
\begin{figure}\begin{center}
\includegraphics[scale=0.51]{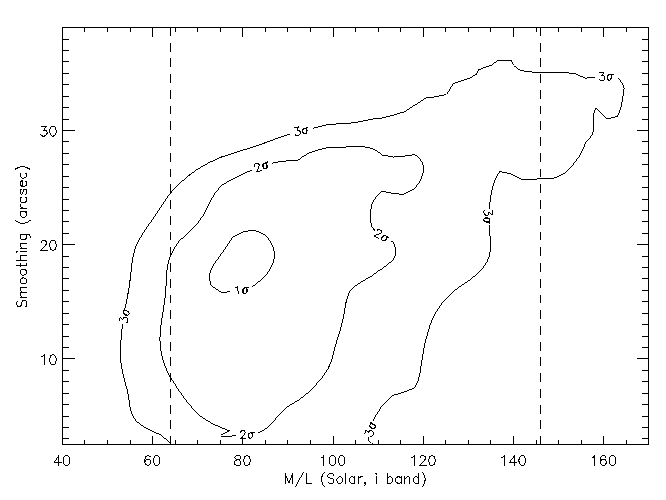}
\caption{Results: constraints on the galaxy group mass-to-light ratio M/L and smoothing scale $s$ that
characterises the size of dark matter clumps.
Vertical dashed lines mark the lower and upper bounds of the
constraint (1\,$\sigma$ error bars) on the group mass-to-light ratio from an independent weak-lensing analysis (Section 7).
}
\label{results}
\end{center}
\end{figure}

$\chi^2$ differences between models with different $s$ and M/L values
are translated into confidence levels, which are 
drawn on Fig.~\ref{results}.
Considering the 2$\sigma$ contour, we find M/L\,=\,98$\pm$27
($i$ band, solar units, not corrected for evolution) and $s$\,=\,20$\arcsec \pm$9.

\subsection{Is Mass Traced by Light ?}
We are able to reproduce accurately the observational constraints when considering an external mass
perturbation drawn from the light distribution. Because our SL analysis is sensitive to the mass, this finding 
is consistent with the hypothesis that light is a good tracer of mass.
We note, however, that we have not demonstrated the uniqueness of the smoothed light model.

\section{Local Effect of Large Scale Perturbation}

In this Section, we propose to explain why a \emph{local} SL analysis is able to constrain \emph{global} 
properties of the galaxy group hosting the lens. First, we investigate the impact of the 
external perturbation on the local SL modelling  (i.e. on the local image positions).

\subsection{Lensing Properties of the External Perturbation}
The lensing properties of a mass distribution are commonly parametrised by a shear $\gamma$ and a 
convergence $\kappa$ \citep[see, \emph{e.g.}][for the definition of these quantities]{bible}.
Here we estimate the mean shear and convergence experienced locally by the lens for each set of
parameters ($s$, M$_{\rm ext}$/L$_{\rm ext}$) by averaging $\gamma$ and $\kappa$ 
generated by the perturbation within a 7$\arcsec$ square encompassing all the multiple images.

\begin{figure*}
\begin{center}
\includegraphics[scale=0.67]{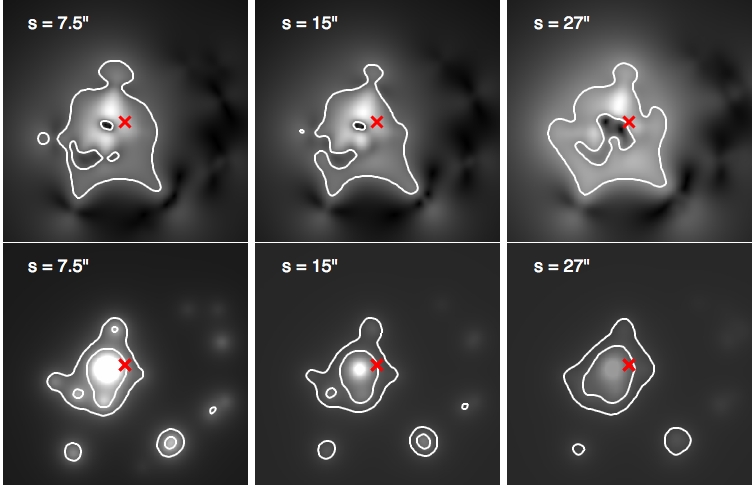}
\caption{Shear (upper panels) and convergence (lower panels) maps generated by the external perturbation 
and experienced by the SL system whose centre is given by the red cross.
The total mass is fixed to the same value (5.7\,10$^{14}$ M$_{\sun}$) in all panels. Panel sizes are
600\,$\times$\,600 square arcseconds, and the smoothing scales $s$ vary as indicated on
each panel.  White contours correspond to shear levels of 0.1 and convergence levels of 0.1 and 0.2.
One can appreciate how the shear and convergence generated by the group are correlated with the smoothing scale.}
\label{Ext_kappa_gamma}
\end{center}
\end{figure*}

Figure~\ref{Ext_kappa_gamma} shows $\kappa$ and $\gamma$ maps generated by an external perturbation
of fixed mass (5.7$\times$10$^{14}$ M$_{\sun}$) for different values of $s$ (reported on each panel).
These maps have been generated for a source redshift of 1.268. Red crosses indicate the lens 
position. One can appreciate how the experienced shear and convergence are correlated with the 
smoothing scale.

\subsection{Shape of the Constraints}

\begin{figure*}
\begin{center}
\includegraphics[scale=0.54]{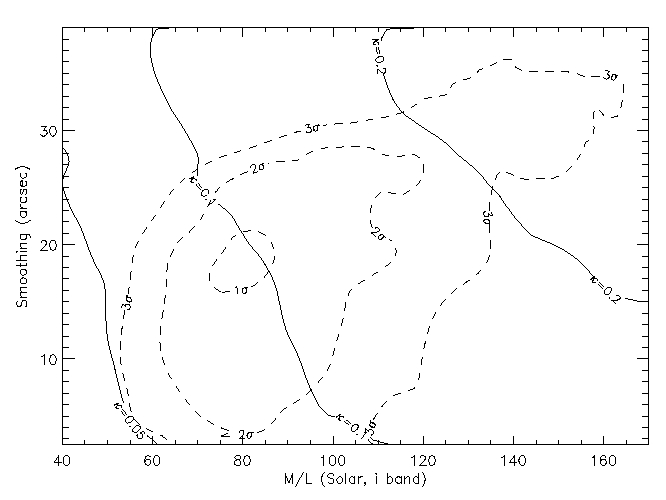}
\includegraphics[scale=0.54]{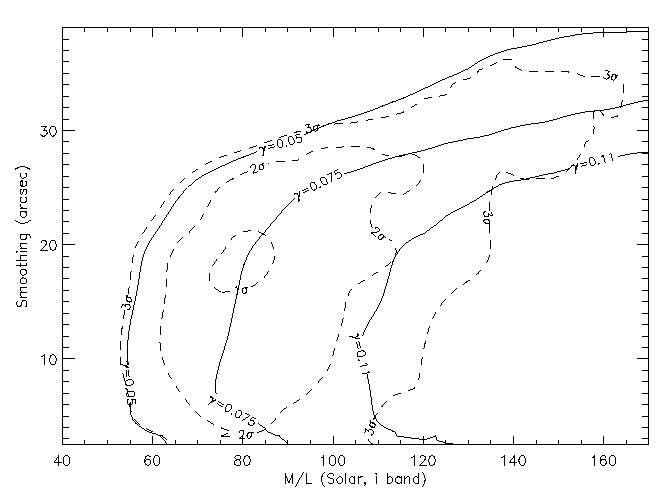}
\caption{
As in Fig.~\ref{results}, constraints obtained on the galaxy group as a whole derived from the \emph{local} 
SL analysis are shown as dashed contours. Solid lines corresponds to lines of constant $\kappa$ (left) 
and $\gamma$ (right) generated by the external mass perturbation and experienced locally by the lens, their
values are labelled on each line.}
\label{results_kappa_gamma}
\end{center}
\end{figure*}

Figure~\ref{results_kappa_gamma} shows contours of $\kappa$ and $\gamma$
overlaid on the results from 
Fig~\ref{results}. We see that the constraints inferred from the SL analysis do not follow $\kappa$ 
contours but do follow $\gamma$ contours. In particular, the best model generates a shear of $\sim$\,0.075.
We interpret the shape of the constraints as follows: one needs to generate a shear value of 
$\sim$\,0.075 with a mass distribution parametrised by a smoothing scale and a total mass.
For a given total mass, the smoother this mass distribution (the higher $s$), the smaller
the generated shear. Therefore, the smoother the mass distribution, the higher the total mass
in order to generate a given shear level.

\subsection{Why Closed Contours ? \\ Shear, Convergence, and Beyond}
Figure~\ref{results_kappa_gamma} suggests that the observational constraints require locally an external 
shear component of $\sim$\,0.075. However, our analysis rejects some external mass perturbations
characterised by such a shear (Fig.~\ref{results}). This suggests, as found by \citet{keeton04}, that 
\emph{the shear approximation fails to capture all of the environmental effects}.
In other words, the shape of the contours for the constraints follows the curves of constant shear. However, the 
contours do close, which means that the constraints are sensitive to more than the shear,
most probably to \emph{higher order terms beyond the shear} that are naturally provided by the modelling
proposed in this work.

In addition, we estimate the shear experienced by images 1.1, 1.2, 1.3 and 1.4. 
To do so, we consider all models falling in the 1$\sigma$ contour. For each model, we compute
the shear experienced by the images, and from these numbers, we estimate the mean shear and the
associated standard deviation. The corresponding shear values for the
images listed above are 0.075, 0.074, 0.073 and 0.073 respectively, 
with a typical
uncertainty of 0.01. Therefore, each image does experience the same shear value within the error bars.
We investigate further the differences between our approach and a constant external shear
approach in the Appendix.

\section{Mass-to-Light Ratio from Weak Lensing}
\label{analysisWL}
We have presented constraints on the mass-to-light ratio of galaxy group \lens\,\, based on a local SL 
analysis. In this Section, we aim to constrain the mass-to-light ratio of \lens\,\, from an independent
weak lensing (WL) analysis, which is intrinsically more sensitive to the projected mass distribution on large 
scales. The goal is to check whether the M/L ratios inferred from the two techniques are consistent.

For a detailed description of our WL methodology see \citet{paperI}. Here we give a brief reminder.
We select as background sources all galaxies with $i$-band magnitudes
in the range 21.5-24. The resulting galaxy number density 
is 13.5 arcmin$^{-2}$. The completeness magnitude in this band is 23.91 and the seeing is $\sim$0.51$\arcsec$.
A Bayesian method, implemented in the \textsc{Im2shape} software \citep{im2shape}, is used to fit the shape
parameters of the faint background galaxies and to correct for the PSF smearing. From the catalogue of 
background galaxies, \citet{paperI} performed a one-dimensional WL analysis. They fit a Singular Isothermal 
Sphere (SIS) model to the reduced shear signal between 150\,kpc and 1.2\,Mpc from the group centre,
finding an Einstein radius of 5.4$\pm$2.1$\arcsec$. In order to relate the strength of the WL signal to a 
physical velocity dispersion characterising the group potential, \citet{paperI} estimate the mean geometrical 
factor using the photometric redshift catalogue from the T0004 release of the CFHTLS-Deep survey
\footnote{http://www.ast.obs-mip.fr/users/roser/CFHTLS\_T0004/} \citep{iena}. They find 
$\sigma_{\rm SIS} = 658^{+119}_{-146}$ km\,s$^{-1}$. This translates into a total mass within the considered 
square of 5.3$\pm$2.0 10$^{14}$ M$_{\sun}$. Because the total luminosity is 5\,10$^{12}$ L$_{\sun}$, we 
find a mass-to-light ratio of 106$\pm$40 ($i$ band, solar units, not corrected for passive evolution).

This is comparable to the M/L constrained by SL \emph{only}. The good agreement (Fig.~\ref{results}) 
between the two methods gives support to the analysis based on SL only.

\section{Discussion}
\label{discuss}
\subsection{Mass is Traced by Light}
An external mass perturbation
derived from the light of the group members allows us to fit accurately the observed constraints.
Because the observed constraints are sensitive to mass rather than light, this suggests that light is a good tracer of 
mass. We note that this result brings forth an efficient way of taking into account an external mass
perturbation in SL modelling. Indeed, this perturbation is fully described with only two parameters, 
the mass-to-light ratio and the smoothing scale. In contrast, describing this perturbation parametrically 
using a mass clump would require at least three parameters (position and velocity dispersion), unless 
independent data constrain one or more of these parameters \citep[see, \emph{e.g.}][where X-ray observations allow
one to constrain the group centre]{hong09}.

\subsection{What is the source of the Constraints ?}
\label{no_image1.4}
Why does our SL analysis allow us to infer properties on the whole galaxy 
group? We claim that this is due to the perturbed state of the SL system of \lens.
Most of the perturbed signal of the multiply imaged system comes from image 1.4, because it is located
farther from the lens centre ($\sim 8\arcsec$) than images 1.1, 1.2 and 1.3 ($\sim 5\arcsec$).
If we remove image 1.4 from the set of observational constraints, we are able to fit very well the remaining 
images without considering any external mass perturbation (the lens being modelled as in Section~3.4.)
In that case, we get RMS\,=\,0.03$\arcsec$ and a reduced $\chi^2$ equals to 0.03.
Therefore, ignoring image 1.4 prevents us from putting any constraints on the external mass perturbation, i.e. the 
host galaxy group. This shows that image 1.4 yields the constraints presented in this work. This finding will 
help us diagnose the type of the SL systems to which our new
analysis technique can be applied (see Section~\ref{lookfor}).

We note that the SL analysis presented here is very simple since we just conjugate a couple of images 
with each other. In particular, we do not use the constraints coming from the whole Einstein ring. More 
sophisticated methods fully take into account arc surface brightness constraints 
\citep[see, \emph{e.g.}][]{warrendye,suyu06,barnabe}. We are aware that we ignore some information that could 
allow us to put stronger constraints on the galaxy group. On the other hand, the basic level of the SL
analysis done here emphasises even more the prospects of this method.

\subsection{The Satellite Galaxy}
We have assumed that the satellite galaxy does not produce significant shear on the images used as 
constraints. However, one could argue that neglecting this satellite galaxy effectively produces the 
claimed constraints from the SL analysis. This is not likely -- due to the location of the satellite galaxy 
with respect to the multiple images (Fig.~\ref{present}), the satellite galaxy may produce a marginal shear on 
on images 1.1, 1.2 and 1.3, 
but is unlikely to have any significant influence on image 1.4,
the image that yields most of the 
constraints. Indeed, the distance between the satellite galaxy and image 1.4 is $\sim$ 13$\arcsec$.
We note that we do not quantify the bias that could result under our working assumption.

Besides, a paper focusing on the properties of the satellite galaxy
(Suyu \& Halkola, submitted to A\&A) shows  
that even with the satellite galaxy included in the lens model, an external shear of roughly 
the same magnitude is needed to fit the observed constraints.

\subsection{Choice of the Lens' Scale Radius}
The dPIE scale radius is where the logarithmic slope of the 3D density profile
smoothly decreases from -2 to -4.
The scale radius of the lens is set to 250\,kpc in the present analysis. 
We have also done a complete analysis for a scale radius of 400\,kpc as a sanity check and 
found that the results inferred for the group do not change significantly.
To understand why, we superimposed critical lines of the best-fit parameters of Table~\ref{res}, for 
a source redshift of 1.268 (without external perturbation), and the critical lines of the best-fit 
parameters of Section~5.1 (with external mass perturbation). 
We find that the external mass perturbation generates a critical line shift of 1.3$\arcsec$.
In parallel, we investigated the critical lines shifts between various scale radii; increasing from 250\,kpc to 400\,kpc and decreasing from 250\,kpc to 100\,kpc. The shifts are 0.12$\arcsec$, an order of magnitude smaller than the shift due to the external mass perturbation.

\subsection{Looking for Perturbed SL Systems}
\label{lookfor}
We propose to characterise the kind of perturbed SL systems one should
target in order to perform 
analyses similar to the one presented in this work. From the ring test done in Abell~1689 
\citep[][see Section~1.3]{ring1689} and the analysis presented in this
paper, we hint at the need for
{\emph{a non-dominant SL system to be used as a test particle for probing the main potential.}}

This is linked to the global geometry of the structure hosting the SL system: to be affected by a perturbation, 
the SL system should not be at the centre of the structure. Indeed, if the lens studied in this paper would have 
dominated the whole group potential, image 1.4 would have been located at a similar distance from the lens 
centre as images 1.1, 1.2 and 1.3. 

The Cosmic Horseshoe \citep{belokurov07,dye08} illustrates this point: 
it is an almost complete Einstein ring of radius 5$\arcsec$ containing a luminous red galaxy in its centre.
As revealed by the SDSS photometry, this galaxy is the brightest object in the group of $\sim$\,26 members  
and it dominates the group light distribution. No external shear is required in the model of the Cosmic 
Horseshoe SL system, which is already suggested by the nearly perfect circle outlined by the ring.

To summarise, we should look for multiply imaged systems where one of the images is found at a
larger radial distance than the other images of the SL system.

\subsection{\lens: Further Evidences for a Bimodal Mass Distribution from
Spectroscopy of Group Members}
We have shown that the modelling based on strong lensing \emph{only} provides strong hints for
a bimodal mass distribution: the first mass component is clearly associated with the strong lensing deflector, and
the second one that perturbs the strong lensing configuration seems,
to first order, to be associated
with the second light peak of the bimodal light map. This suggests a dynamically young structure in the
process of formation. A spectroscopic survey of the group further
supports this hypothesis: 
we measured redshifts for 36 galaxies along the direction of \lens\,\,by using spectroscopic data acquired with FORS2 at the ESO Very Large Telescope (VLT), and confirmed the presence of a high concentration of galaxies at $z\sim0.35$ (Mu\~noz et~al.,
in preparation). A careful analysis of the redshift distribution of
galaxies around this peak reveals two close structures with a radial
velocity difference of $V_r=1\,180\,{\rm{km\,s}}^{-1}$.
This result is in agreement with the interpretation of our strong lensing \emph{only} analysis.

\section{Conclusion}

We propose a method to constrain the dark matter distribution of galaxy groups and clusters.
Exploiting information contained in perturbed SL systems, we use the SL geometry to probe the main 
potential of the host structure responsible for that perturbed state.

We show that the SL \emph{only} constraints on the mass-to-light ratio of \lens\, are in good agreement 
with WL constraints obtained independently, supporting the reliability of the proposed method. 
Moreover, the SL \emph{only} analysis provides strong hints for a bimodal mass distribution, which is confirmed
by the spectroscopic survey of galaxy group members.

We advocate the need for a dedicated search of perturbed SL systems in the HST archive in order to test and 
validate further this method, which is particularly promising in the light of future large surveys that will yield thousands 
of SL systems, some of them being perturbed enough to perform the test presented in this paper.

\section*{Acknowledgement}
ML acknowledges Bernard Fort, Masamune Oguri \& Phil Marshall for related discussions.
ML acknowledges the anonymous referee for a detailed report, and Christopher Kochanek for insightful 
comments on the submitted version of this paper. ML acknowledges the Centre National d'Etudes Spatiales 
(CNES) and the Centre National de la Recherche Scientifique (CNRS) for their support. ML est b\'en\'eficiaire 
d'une bourse d'accueil de la Ville de Marseille. The Dark Cosmology Centre is funded by the Danish National 
Research Foundation. We thank the Danish Centre for Scientific Computing at the University of Copenhagen 
for providing a generous amount of time on its supercomputing facility. 
EJ is supported by the NPP, administered by Oak Ridge Associated
Universities through a contract with NASA. Part of this work was carried out
at Jet Propulsion Laboratories, California Institute of Technology under a
contract with NASA.
JR acknowledges support from an EU 
Marie-Curie fellowship. SHS is supported in part through the Deutsche Forschungsgemeinschaft (DFG) under 
project SCHN 342/7-1, and AH by the DFG cluster of excellence 'Origin and Structure of the Universe'.
JPK acknowledges CNRS for its support.

\bibliographystyle{aa} 
\bibliography{draft}

\newpage
\appendix
\section{Taking an External Mass Perturbation Into Account: Comparison with Other Approaches}

We have proposed in this article a way of taking into account an external mass perturbation in a strong 
lensing (SL) modelling. Here we try other possible and more conventional approaches: i) a constant external 
shear profile and ii) a Singular Isothermal Sphere centred on the second high luminosity peak, which, by
construction, is the main mass concentration perturbing the SL in the method proposed in this work.

\subsection{A Constant External Shear}
Although unphysical (any mass distribution will not generate shear only but also convergence), the external
shear model is widely used and is often a good approximation. Here we address the modelling of the 
SL system with a constant external shear component parametrised by a position angle and a strength 
($\gamma_{\rm Kst}$). This modelling is performed in the image plane. Parameters of the potential describing 
the lens are set as in Section~\ref{singlemodelling}. The external shear strength is allowed to vary between 
0 and 0.3. The upper limit corresponds to a very strong shear value: for comparison, the massive galaxy 
cluster Abell~1689 produces an average shear value of 0.23 at 50$\arcsec$ away from its 
centre.

We are able to get a very good fit, with $\chi^2 < 1$.
The best model corresponds to a circular halo for the lens ($e=0.036$) centred on the bright galaxy
(X\,=\,-0.36$\arcsec$, Y\,=\,0.07$\arcsec$), making its position angle (95 degrees) irrelevant.
The lens fiducial velocity dispersion equals to 450$\pm$8 km\,s$^{-1}$ (1$\sigma$).
The external shear is described by $\gamma_{\rm Kst}$\,=\,0.19, more than twice the one
derived in Section 6.2 ($\sim$\,0.075), and a position angle equal to 19.8 degrees.

The \textsc{Lenstool} software does explore the parameter space using a MCMC
sampler \citep{jullo07}. Therefore, we can use these MCMC realisations in order to investigate the
degeneracies between the different parameters. The following figures have been generated
this way.

Figure~\ref{degen1} shows that there is a strong degeneracy between $e$
and $\gamma_{\rm Kst}$. We see that the solution derived in Section~5.1 (i.e. an external shear of
$\sim$\,0.075 and an ellipticity of $\sim$\,0.5) is included in the 1-$\sigma$ contour.
On the other hand, the position angle of the external shear is very well constrained to be
$\sim$\,20 degrees. This position angle points towards the second high luminosity light clump. 
This suggests that, to first order, the external mass 
perturbation is dominated by this component.
We note that the best-fit model needs an external shear of order 0.18, which is a pretty unlikely value 
in our case because it is comparable to what would be experienced at $\sim$ 100$\arcsec$ from the centre of Abell~1689.

\begin{figure}[h!]
\begin{center}
\includegraphics[scale=0.5]{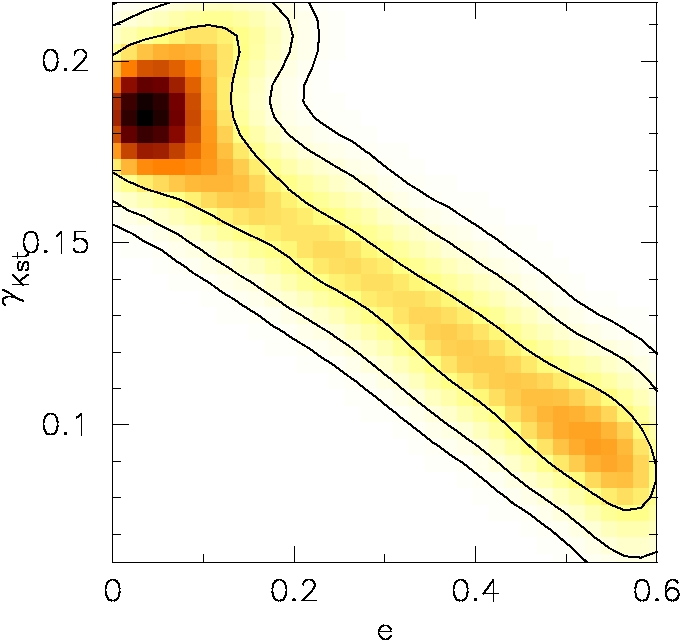}
\includegraphics[scale=0.55]{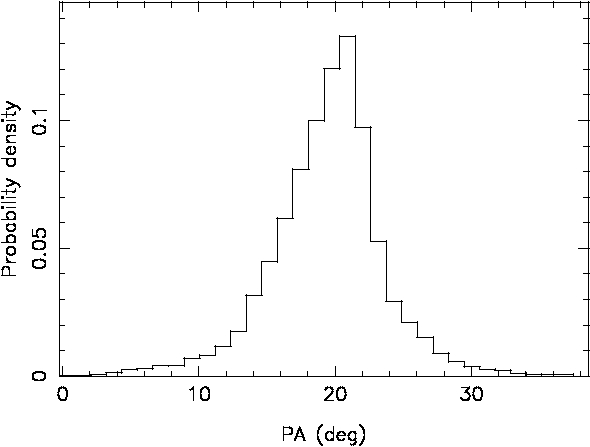}
\caption{
Results of a constant external shear model. \emph{Top:} degeneracy between the halo ellipticity $e$ and 
the strength of the external shear $\gamma_{\rm Kst}$. 
\emph{Bottom:} posterior probability distribution for the position angle of the external shear.
}
\label{degen1}
\end{center}
\end{figure}

\subsection{An SIS profile}
The first order mass perturbation is associated to a second peak of high luminosity. We put an SIS mass 
distribution at the location of this second luminosity peak (X,Y\,=\,-53,10$\arcsec$ wrt the lens). We allow its 
velocity dispersion to vary up to 800 km\,s$^{-1}$, an upper limit motivated by the WL analysis of the full 
group (Section~7), and do the SL modelling with parameters set as in Section~\ref{singlemodelling}. We are 
able to get a very good fit, with $\chi^2 < 1$. The lens halo is centred on the 
bright galaxy. Its ellipticity equals 0.43$^{+0.01}_{-0.12}$ and its position angle  27$\pm$2 deg. 
The lens fiducial velocity dispersion equals to 
441$\pm$7 km\,s$^{-1}$ (1$\sigma$). The external shear and convergence generated by the SIS profile at 
the location of the multiple images are equal by definition, between 0.07 and 0.10.

\begin{figure}[h!]
\begin{center}
\includegraphics[scale=0.5]{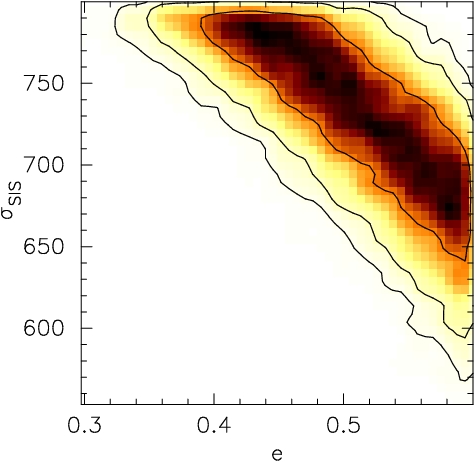}
\caption{Degeneracies between the lens halo ellipticity and the SIS profile velocity dispersion, related to the 
strength of the external shear experienced by the multiple images.}
\label{degen2}
\end{center}
\end{figure}

We show on Fig.~\ref{degen2} the degeneracies between the  lens halo ellipticity and the SIS profile velocity 
dispersion, related to the strength of the external shear experienced by the multiple images.
We see that the solution derived in Section~5.1 (i.e. an external shear of $\sim$0.075, corresponding to 
$\sigma_{\rm SIS} \sim$\,700\,km\,s$^{-1}$ and an ellipticity  of $\sim$0.5) is included in the 1-$\sigma$ 
contour.

\subsection{Discussion}
In each cases investigated in this Appendix, we find that the solution we have derived  in Section~5.1 using 
our original method is consistent with solutions derived with more conventional methods.

We note that conventional methods exhibit strong degeneracies between the lens halo ellipticity and the 
strength of the external shear. These degeneracies are smaller in the case of the SIS profile (lens ellipticity is
constrained between 0.3 and 0.6) compared to the case of a constant shear profile (ellipticity unconstrained 
between the allowed priors: 0 and 0.6). The main difference between the SIS profile and the constant shear 
profile is that the SIS profile generates both shear and convergence.

With respect to the lens itself, we note that all fitted fiducial velocity dispersions are consistent, whatever the 
method used to take into account the external mass perturbation. They fall between 433 km\,s$^{-1}$ and 
458 km\,s$^{-1}$. This translates into a projected mass computed in a radius of 10$\arcsec$ between 
0.93 and 1.04$\times$10$^{13}$ M$_{\sun}$. This is expected because the mass of the lens within this radius 
is set by the location of the SL constraints and therefore does not depend much on the external mass 
perturbation (see also Section~4.3).

\end{document}